\documentclass[aps,prl,twocolumn,groupedaddress]{revtex4-1}
\usepackage{graphicx}
\usepackage{amsmath}
\usepackage{amssymb}
\usepackage{graphicx}
\usepackage{dcolumn}
\usepackage{bm}
\usepackage{color}
\newcommand{\ypp}{\rm Yb$_{2}$Pt$_{2}$Pb}
\newcommand{\scbo}{SrCu$_{2}$(BO$_{3}$)$_{2}$}

\begin{document}






\title{Magnetic structure of Yb$_{2}$Pt$_{2}$Pb: Ising moments on the Shastry-Sutherland Lattice}
\author{W. Miiller$^1$, L.S. Wu$^1$, M. S. Kim$^1$, T. Orvis$^1$, J. W. Simonson$^1$, M. Gam\.{z}a$^2$, D. E. McNally$^1$, C. S. Nelson$^3$, G. Ehlers$^4$, A. Podlesnyak$^4$, J.S. Helton$^5$, Y. Zhao$^{5,6}$, Y. Qiu$^{5,6}$, J. R. D. Copley$^5$, J. W. Lynn$^5$, I. Zaliznyak$^{2}$, and M. C. Aronson$^{1,2}$}
\email[]{maronson@bnl.gov}
\affiliation{$^1$ Department of Physics and Astronomy, Stony Brook University, Stony Brook, NY 11794}
\affiliation{$^2$ Condensed Matter Physics and Materials Science Department, Brookhaven National Laboratory, Upton, NY 11973}
\affiliation{$^3$ National Synchrotron Light Source, Brookhaven National Laboratory, Upton, NY 11973}
\affiliation{$^4$ Quantum Condensed Matter Division, Oak Ridge National Laboratory, Oak Ridge, Tennessee 37831}
\affiliation{$^5$ NIST Center for Neutron Research, Gaithersburg, MD 20899}
\affiliation{$^6$ Department of Materials Science and Engineering, University of Maryland, College Park, MD 20742}
\date{\today}

\begin{abstract}
Neutron diffraction measurements were carried out on single crystals and powders of ~\ypp, where Yb moments form planes of orthogonal dimers in the frustrated Shastry-Sutherland Lattice (SSL). ~\ypp~ orders antiferromagnetically at T$_{N}$=2.07 K, and the magnetic structure determined from these measurements features the interleaving of two orthogonal sublattices into a 5$\times$5$\times$1 magnetic supercell that is based on stripes with moments perpendicular to the dimer bonds, which are along (110) and (-110). Magnetic fields applied along (110) or (-110) suppress the antiferromagnetic peaks from an individual sublattice, but leave the orthogonal sublattice unaffected, evidence for the Ising character of the Yb moments in ~\ypp. Specific heat, magnetic susceptibility, and electrical resistivity measurements concur with neutron elastic scattering results that the longitudinal critical fluctuations are gapped with $\Delta$E$\simeq$0.07 meV.

\end{abstract}


\maketitle


Much attention has focussed on coupled spin-dimer systems, where frustrated magnetic order can lead to spin liquid states~\cite{rice2002,balents2010}. The Shastry-Sutherland Lattice (SSL), consisting of planes of orthogonal dimers with intradimer exchange $J$ and interdimer exchange $J^{\prime}$, is of particular interest as it has been  solved~\cite{shastry1981} to show that the ground state is ordered antiferromagnetically (AF) for $J/J^{\prime}\leq (J/J^{\prime})_{C}$ and otherwise has a singlet ground state with gapped magnetic excitations. Insulating ~\scbo~ has such a  SSL dimer liquid ground state~\cite{miyahara2003,gaulin2004,kakurai2005}, where frustrated exchange interactions create complex ordered superstructures at high fields~\cite{momoi2000,kodama2002,takigawa2013}, resulting in quantized plateaux in the magnetic field H dependence of the magnetization M(H)\cite{sebastian2008,jaime2012,matsuda2013}.

Most SSL systems are magnetically ordered, where the closure of the singlet dimer gap yields a nonzero M(H) that makes AF order possible even for H=0 ~\cite{michimura2006,siemensmeyer2008,yoshii2008,matas2010,kim2008}. In metallic ~\ypp, Yb moments form the SSL (Fig.~1a) and AF order appears below  T$_{N}$=2.07 K~\cite{kim2008}. Magnetic susceptibility $\chi$(T) measurements~\cite{kim2013} find that J$\simeq$5 K and  $J/J^{\prime} \simeq$1, and a field of only 1.23 T, applied along the (110) or (-110) dimer bonds, suppresses the H=0 AF order, accompanied by quantized M(H) steps ~\cite{shimura2012,iwakawa2012,kim2013} that culminate in the saturation of M(H) at $\simeq$3 T. The magnetic structures of SSL systems reflect the near-frustration of short-ranged magnetic interactions, and in ~\ypp, where strong Ising anisotropy confines large and classical Yb moments to the SSL planes, the H=0 AF structure is analogous to but distinct from those found in ~\scbo~ at high fields, where quantum Heisenberg  spins S=1/2 are perpendicular to the SSL layers.

Neutron diffraction measurements presented here demonstrate that the Ising Yb moments in ~\ypp~ have a  remarkably complex AF structure, with a commensurate superlattice of ordered stripes consisting of moment-bearing and magnetically-compensated Yb-pairs with moments perpendicular to the (110) and (-110) dimer bonds. The magnetic cell contains 200 Yb atoms, arranged in two orthogonal sublattices (SLs) that can be separately polarized by magnetic fields. The persistence of AF order in the surviving SL shows that the two orthogonal bonds of the H=0 SSL are decoupled by field, leading to a unique high field ground state. The AF order parameter in ~\ypp~ is mirrored in the temperature dependencies of the specific heat C/T and the temperature derivatives of $\chi$(T) and the electrical resistivity $\rho$(T), suggesting that the stripe-ordered ground state is preceded by the dynamical freezing of longitudinal critical fluctuations that are gapped.

Crystals of \ypp~ were grown from Pb flux \cite{kim2008}. Neutron scattering experiments were performed using an incident neutron wavelength of 2.359 $\AA$ on the BT-7 triple axis spectrometer at the NIST Center for Neutron Research (NCNR)~\cite{lynn2012} on 5 g of powder prepared by triturating single crystals, and on a 60 mg single crystal. x-ray diffraction measurements were carried out on this same crystal on beamline X21 at the National Synchrotron Light Source (NSLS), with an incident x-ray energy of 10 keV. Neutron scattering experiments were conducted on a 6 g collection of $\sim$ 300 oriented single crystals of \ypp with an angular mosaic perpendicular to the c-axis od $\simeq$ 2 degrees, using an incident energy E$_{i}$=3.27 meV at the Disk Chopper Spectrometer (DCS) at NCNR~\cite{copley2003}, while E$_{i}$=3.32 meV was used at the Cold Neutron Chopper Spectrometer (CNCS) at the Spallation Neutron Source.  Measurements of $\chi$(T) were carried out using a Quantum Design Magnetic Phenomena Measurement System, while $\rho$(T) and C were measured using a Quantum Design Physical Phenomena Measurement System.

Neutron diffraction measurements show clear evidence for AF order below T$_{N}$=2.07 K ~\cite{kim2008}. Fig.~1a shows the elastic intensity in the (HHL) scattering plane at 1.5 K, obtained at CNCS on the collection of aligned ~\ypp~ crystals. The (002) and (112) nuclear peaks are present at all temperatures, however satellite peaks emerge around  (000) and  (111) for $T\leq T_{N}$.  The AF wave vector $\bf{q_{1}}=(\delta, \delta, 0)$ with $\delta=0.2\pm0.02$ in reciprocal lattice units (rlu) is deduced from Gaussian fits to the (1-$\delta$,1-$\delta$,1) satellites(Fig.~1b), indicating that AF order involves a $5\times 5\times 1$ superstructure. The temperature dependencies of the integrated intensities of the  ($\delta$,$\delta$,1) and (1-$\delta$,1-$\delta$,1) AF peaks have the appearance of order parameters that are substantially broadened near T$_{N}$~(Fig.~1b,inset).

Our scattering geometry accesses a strip of the (HK1) plane (Fig.~1c), where a quartet of AF satellites forms around both (001) and (111), signalling the SSL plane retains fourfold symmetry in AF ~\ypp. Thus, there are two AF wave vectors, $\bf{q_{1}}$=(0.2,0.2,0) and $\bf{q_{2}}$=(-0.2,0.2,0) that define two orthogonal SLs, each containing half of the Yb moments. The fourth (1+$\delta$,1+$\delta$,1) satellite around (111) is much weaker than can be explained by either the Yb magnetic form factor or self-absorption corrections, and we will show below that this is captured by our magnetic structure model. A 3 T field H$\|$(-110) saturates all antiparallel moments~\cite{kim2008,
ochiai2011}, and Fig.~1d shows that only the AF peaks with $\bf{q_{1}} \perp$ H survive.  Their Ising character restricts the Yb moments to be along or perpendicular to the (110) and (-110) dimer bonds. Both configurations are allowed by symmetry, but our magnetic structure produces no AF peaks for moments oriented along the dimer bonds. We conclude that the Yb moments are perpendicular to their own dimer bonds, so the $\bf{q_{1}}$=(0.2,0.2,0) SL is formed by dimers oriented along (110) with  moments along (-110) and for the orthogonal $\bf{q_{2}}=$(-0.2,0.2,0) SL the dimers are along (-110) with moments along (110). The Ising character of the Yb moments is accommodated in ~\ypp~ by the formation of two orthogonal SLs, as the magnetization anisotropy suggests~\cite{ochiai2011}.

\begin{figure} [t]
\includegraphics[width=9cm]{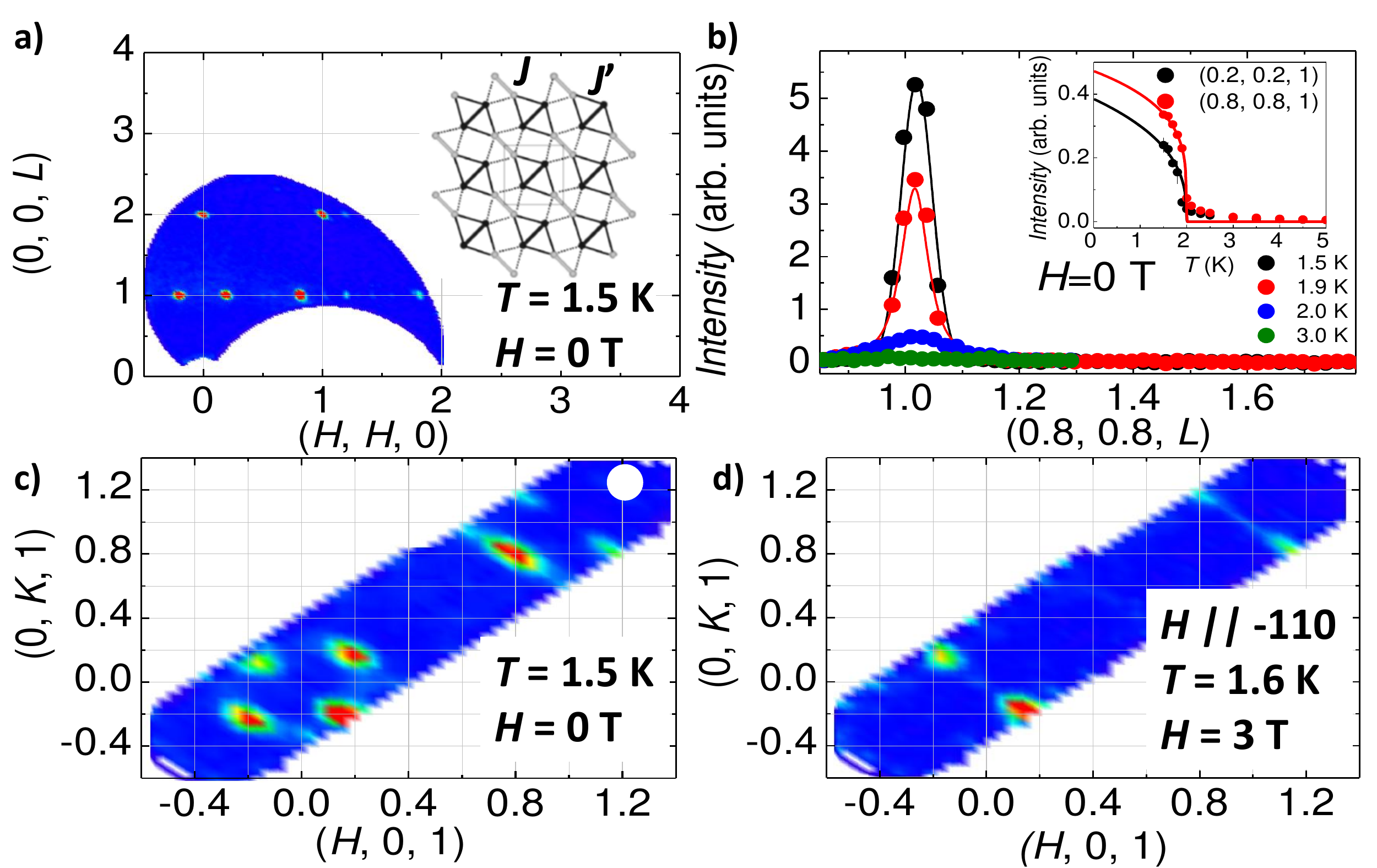}%
\caption{\label{fig_1} a) Elastic scattering in HHL plane at 1.5 K. Inset: orthogonal Yb dimers along (110) and (-110) form the SSL. b) Intensity of the (0.8,0.8,1) AF peak at different temperatures. Solid lines are Gaussian fits. Inset: temperature dependencies of the intensities of the (0.2,0.2,1) (black points) and (0.8,0.8,1) (red points) AF peaks. Solid lines are guides for the eye.  Elastic scattering at 1.5 K in the (HK1) planes for magnetic fields H=0 (c) and at 1.6 K for H=3 T (d).}
\end{figure}

The magnetic structure of ~\ypp~ is obtained from the refinement of neutron diffraction data obtained on a 5g powder using BT-7. The 10 K neutron powder pattern (Fig.~2a) is well refined using the reported U$_{2}$Pt$_{2}$Sn -type tetragonal structure~\cite{pottgen1999}. The 0.5 K neutron powder diffraction pattern has additional peaks corresponding to AF order (Fig.~2a), also evident in the (HH0) scan obtained at CNCS on the aligned crystal array (Fig.~2b). These (1$\pm\delta$,1$\pm\delta$,0) AF satellites arise from Yb moments that are perpendicular to (HH0) and to  $\bf{q_{1}}$. No scans were performed along directions parallel to (-HH0), however we have added to Fig.~2b the second set of AF satellites from the moments perpendicular to $\bf{q_{2}}$. The 1.5 K (HH0) neutron scan (Fig.~2c) carried out on BT-7 using a 60 mg single crystal shows  two satellites  flanking each of the (110), (220), and (330) nuclear Bragg peaks, while their absence in x-ray scans carried out on the same crystal (Fig.~2c) indicates no structural distortion occurs at T$_{N}$.

\begin{figure} [t]
\includegraphics[width=9cm]{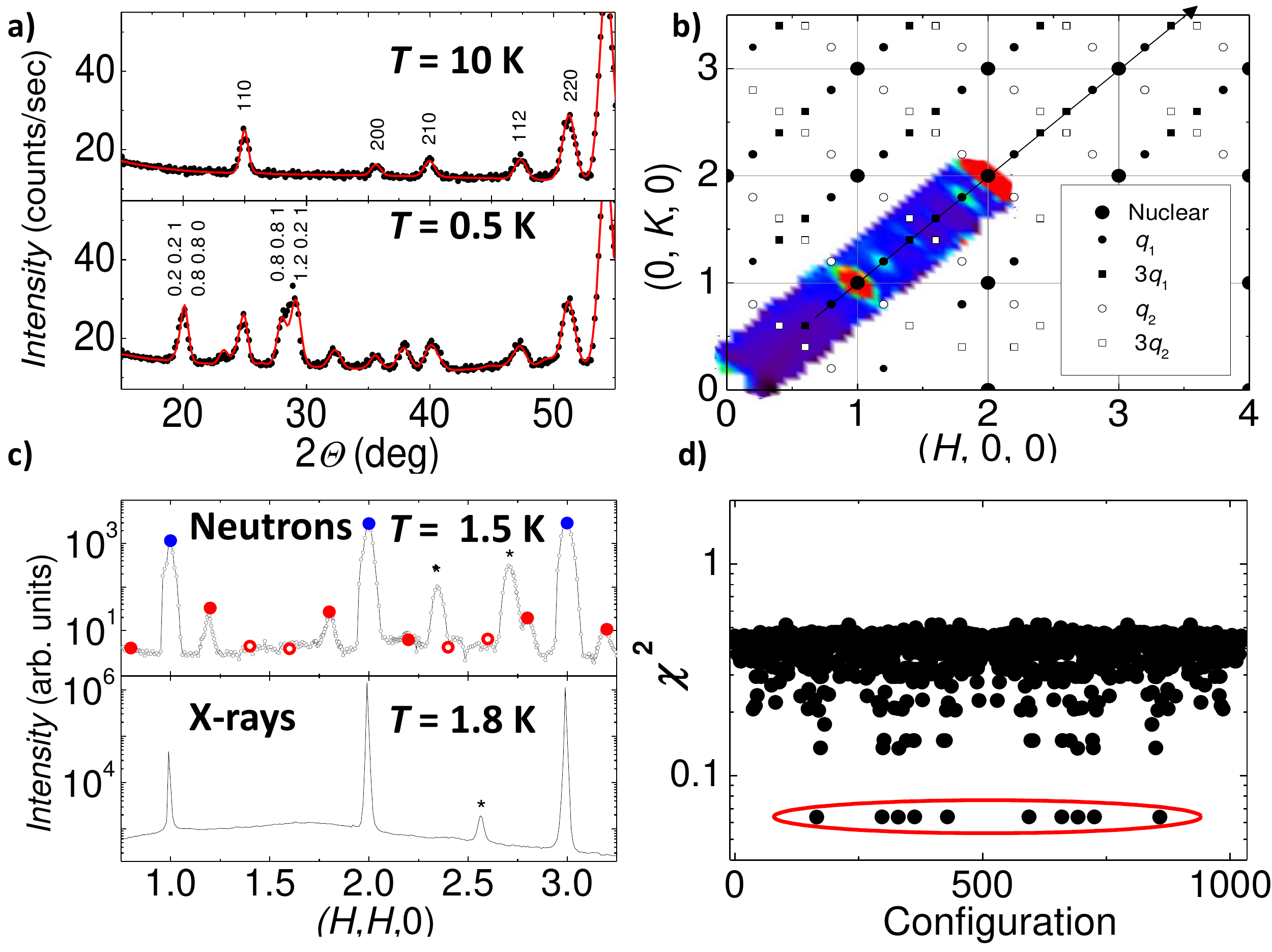}%
\caption{\label{fig_2} a) Neutron powder diffraction patterns at 10 K (top) and 0.5 K (bottom), with indexed AF and nuclear peaks. Red lines are Rietveld fits (see text). b)  HK0 plane for T$\leq$T$_{N}$=2.07 K, indicating nuclear {\Large{$\bf{\bullet}$}} and main AF satellites $\bf{q_{1}}$($\bullet$) and $\bf{q_{2}}$ ($\circ$), as well as AF harmonics 3$\bf{q_{1}}$ ($\blacksquare$) and 3$\bf{q_{2}}$ ($\square$). 1.5 K CNCS data are overplotted. c) Neutron (BT-7, 1.5 K, top) and x-ray (X21, 1.8 K, bottom) HH0 scans of 0.06 g single crystal, with sample holder peaks(stars), nuclear peaks (Filled blue points), and first order (first harmonic) AF peaks (full and open red points)from our magnetic model. d) residual~$\chi^{2}$ for 1024 stripe configurations. Red circle encloses the 10 solutions with lowest $\chi^{2}$=0.064.}
\end{figure}

Since the magnetic structure of \ypp~ is quite complex, involving two different AF wave vectors and the formation of a 5$\times$5$\times$1 supercell with 200 Yb moments, it cannot be determined solely from conventional representation analysis. However, the Ising character of the Yb moments greatly constrains our magnetic model. The anisotropy of M(H) indicates that the Yb moments lie in the SSL planes, and have no components perpendicular to either (110) or (-110). No canting and no  modulation of the Yb moment amplitude is possible. The crystal structure consists of Yb dimers in the SSL planes with long bonds (3.889 $\AA$)  and short bonds(3.545 $\AA$), potentially having different Yb moments M1 and M2. The unit cell contains $z=0$ and $z=1/2$ planes with both types of dimers, but they are staggered so that along the c-axis a long dimer with Yb moment $M1$ ($z=0$) is always below a short dimer with Yb moment $M2$ ($z=1/2$).

 Yb dimers are the building blocks of the magnetic structure. Since the Yb moments of the crystal-field doublet ground state are large and classical, we consider here the four configurations of two distinguishable moments (Fig.~3), i.e. $\uparrow\downarrow$, $\downarrow\uparrow$, $\uparrow\uparrow$, and $\downarrow\downarrow$, and not the singlet-triplet states of two indistinguishable quantum spins as in ~\scbo. The fourfold symmetry of the SSL plane requires two interleaving and orthogonal magnetic SLs in each SSL plane (Figs.~3a,b). Inspired by the stripelike structures of high field ~\scbo~\cite{kodama2002,momoi2000,miyahara2000}, we construct the 5$\times$5$\times$1 AF superlattice of ~\ypp~ from stripes, each with single configurations of Yb pairs having dimer bonds along (110) and moments parallel to (-110) for the SL with $\bf{q_{1}}$ = (0.2,0.2.0) and with dimer bonds along (-110) and moments parallel to (110) for the perpendicular SL with $\bf{q_{2}}$=(-0.2,0.2.0) (Fig.~3). Each SSL plane consists of two perpendicular arrangements of dimer stripes, and both orthogonal SLs are present in the $z=0$ and $z=1/2$ planes (Figs.~3a,b). The AF wave vectors $\bf{q_{1,2}}$=($\pm$ 0.2,0.2,0) indicate there must be a net AF alignment of the neighboring $z=0$ and $z=1/2$ planes, where a moment bearing $\uparrow\uparrow$ dimer stripe in the $z=0$ plane (Fig.~3a), requires an antiparallel moment bearing $\downarrow\downarrow$  dimer stripe in the neighboring $z=1/2$ plane (Fig.~3b). Similarly, stripes of  $\uparrow\downarrow$ dimers in the $z=0$ plane must be stacked beneath stripes of $\downarrow\uparrow$ dimers in the $z=1/2$ plane. For both SLs and in both $z=0$ and $z=1/2$ planes, stripes based on each of the four dimer configurations can be arranged in a total of 1024 different ways within the $5\times 5\times 1$ AF supercell of \ypp.

\begin{figure} [t]
\includegraphics[width=9cm]{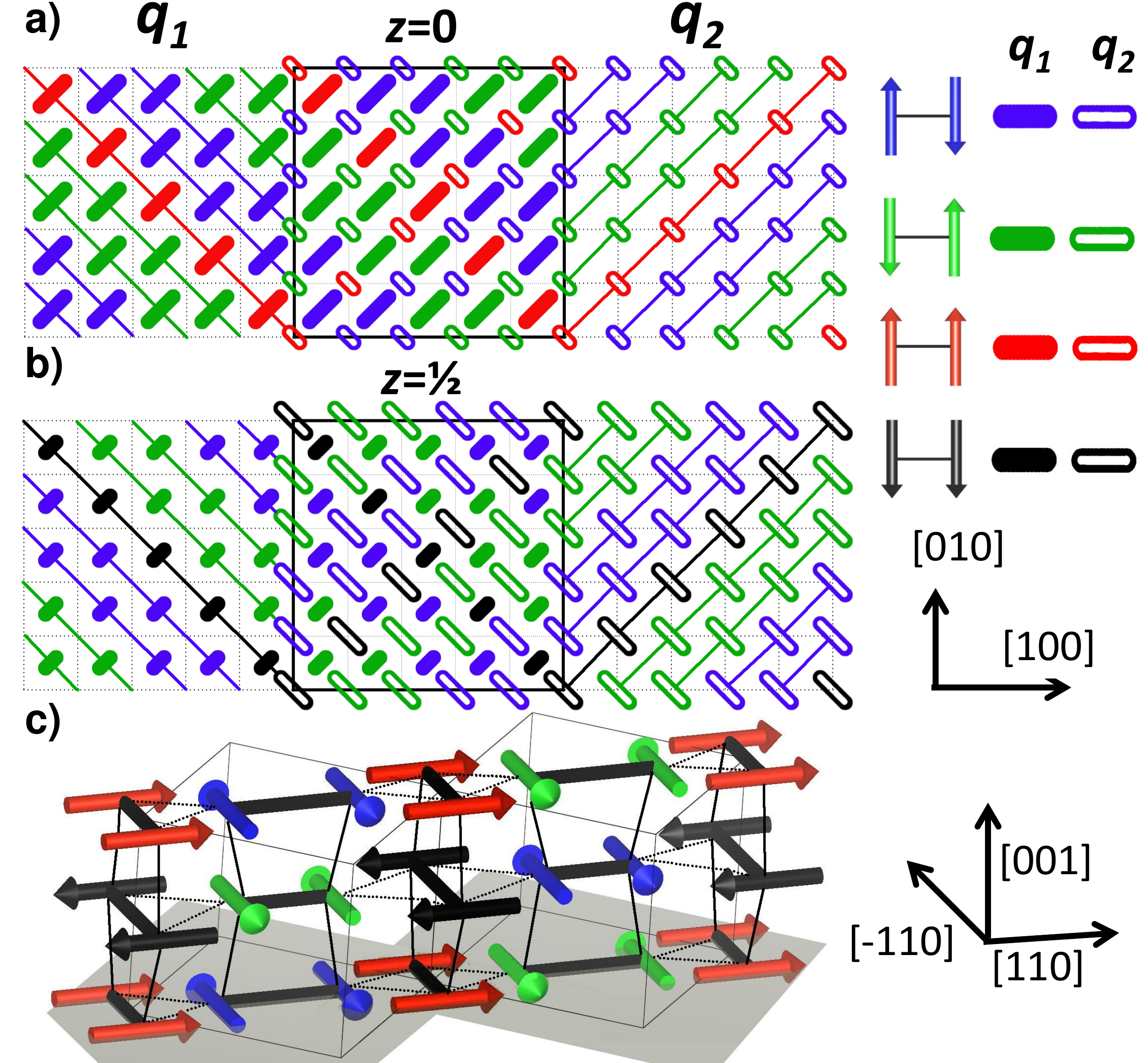}%
\caption{\label{fig_3} Key: the four configurations of classical Yb dimers, on the $\bf{q_{1}}$ (filled symbols) and $\bf{q_{2}}$ (open symbols) sublattices. Symbols are elongated in the direction of the dimer bond, larger (smaller) symbol sizes indicate long (short) dimers. The z=0 (a) and z=1/2 (b) planes are formed by interleaving orthogonal $\bf{q_{1}}$ (left) and $\bf{q_{2}}$ (right) sublattices, to form the 5$\times$5$\times$1 AF supercell (solid lines, center panels). Each sublattice consists of ordered sequences of stripes consisting of single dimer types (different colors). c) A fragment of the complete magnetic structure, showing AF stacking of short and long dimers along the c-axis.}
\end{figure}

These 1024 stripe configurations were refined against the form factor corrected 0.5 K neutron powder data(Fig.~2a) using Fullprof~\cite{rodriguez2001}. The computed diffraction patterns were subtracted from the 0.5 K diffraction pattern to generate a residual $\chi^{2}$ for each configuration. Fig.~2d shows that there are 10 configurations that have markedly smaller values of $\chi^{2}$=0.064, consisting of the five permutations of the stripe sequence ($\uparrow\downarrow$,$\uparrow\downarrow$,$\downarrow\uparrow$,$\downarrow\uparrow$,$\uparrow\uparrow$) and its mirror reflection ($\downarrow\uparrow$,$\downarrow\uparrow$,$\uparrow\downarrow$,$\uparrow\downarrow$,$\downarrow\downarrow$)(Figs.~3a,b). The fit gives the same Yb moment on every site, $M1=M2$ = 3.8 $\pm$0.2$\mu_{B}$/Yb, in good agreement with the 4 $\mu_{B}$/Yb of the Yb$^{3+}$ $|\pm$7/2$>$ Kramers doublet ground state~\cite{aronson2010}, and with the measured saturation moment M$_{110}$~\cite{kim2008,ochiai2011}. The simulated powder pattern that corresponds to the configuration with the lowest $\chi^{2}$ agrees very well with the 0.5 K powder pattern (Fig.~2a). Computed nuclear and AF peak intensities  are also in excellent agreement with the peak heights found in the 0.5 K (HH0) single crystal scan (Fig.~2c).

The refined AF structure for ~\ypp~ is remarkably intricate (Fig.~3). Both SLs in each SSL layer are net-moment bearing, containing individual $\uparrow\uparrow$ or $\downarrow\downarrow$ stripes, spatially separated by two pairs of magnetically compensated $\downarrow\uparrow$ or $\uparrow\downarrow$ stripes for a total of five stripes per magnetic unit cell. The moment-bearing stripes in neighboring SSL layers are arranged AF along the c-axis, and a fragment of the 5$\times$5$\times$1 magnetic unit cell is depicted in Fig.~3c, emphasizing the three-dimensionality of the magnetic structure of ~\ypp. The Ising character of the Yb moments requires two orthogonal SLs in each SSL layer, where half of the moments are parallel to (110) and the other half to (-110). Consequently, only the $\bf{q_{1}}$ SL, which has dimer bonds along (110) and  moments perpendicular to (110), will be saturated by fields in the (-110) direction. The AF peaks associated with the $\bf{q_{2}}$ SL are unaffected by field (Fig.~1d), as is its phase line T$_{N}$(B)=2.07 K~\cite{ochiai2011}, demonstrating that the two SLs are independent.

The directions but not the magnitudes of the Yb dimer moments are modulated by a square wave with a periodicity of 5 chemical unit cells. We observed 3$\delta$ harmonics in the (HH1) scans  of the 6 g array of aligned crystals at 1.5 K (CNCS) and at 0.1 K (DCS) (Fig.~4a). The ratio of the ($\delta$,$\delta$,1) AF peak to the (3$\delta$,3$\delta$,1) harmonic is 0.122 in our model, agreeing with the experimental data at 0.1 K. The model correctly estimates the intensities of the nuclear and main AF peaks at 1.5 K, although the 3$\delta$ harmonics are much weaker than predicted, implying the presence of longitudinal critical fluctuations that are quenched between 1.5 K and 0.1 K. C/T and the temperature derivatives of  $\rho$(T) and  $\chi$(T) (Fig.~4b) all show peaks at T$_{N}$=2.07 K, as well as an additional broad feature at $\simeq$0.8 K.  $\rho$(T) was measured with the current along the c-axis and with a magnetic field oriented along (110), and it is plotted with its  temperature derivative $\partial\rho$/$\partial$T in Fig.~4c for H=0, 1T, and 5 T. The ordering anomaly remains fixed at T$_{N}$=2.07 K, indicating that the $\bf{q_{1}}$ SL continues to affect $\rho$(T), even when the $\bf{q_{2}}$ SL has been fully polarized by the 5 T field. While the general appearance of the broad maximum in $\partial\rho$/$\partial$T changes slightly in field, it remains centered at 0.8 K in all fields.

The picture that emerges from the data in Fig.~4 is that individual Yb dimers form above T$_{N}$, consistent with the intradimer exchange J$\simeq$0.6 meV deduced from fits to $\chi$(T)~\cite{kim2013}. As T$\rightarrow$T$_{N}$, these dimers assemble into a striped magnetic structure, accompanied by the gradual suppression of critical fluctuations in favor of static order. These longitudinal fluctuations freeze between 1.5 K and 0.1 K and the much stronger temperature dependencies in C/T and the temperature derivatives of $\chi$(T) and $\rho$(T) above 0.8 K are together indications that these critical modes are gapped by $\Delta$E/k$_{B}\simeq 0.8$ K, where k$_{B}$ is Boltzmann's constant.

\begin{figure} [t]
\includegraphics[width=9cm]{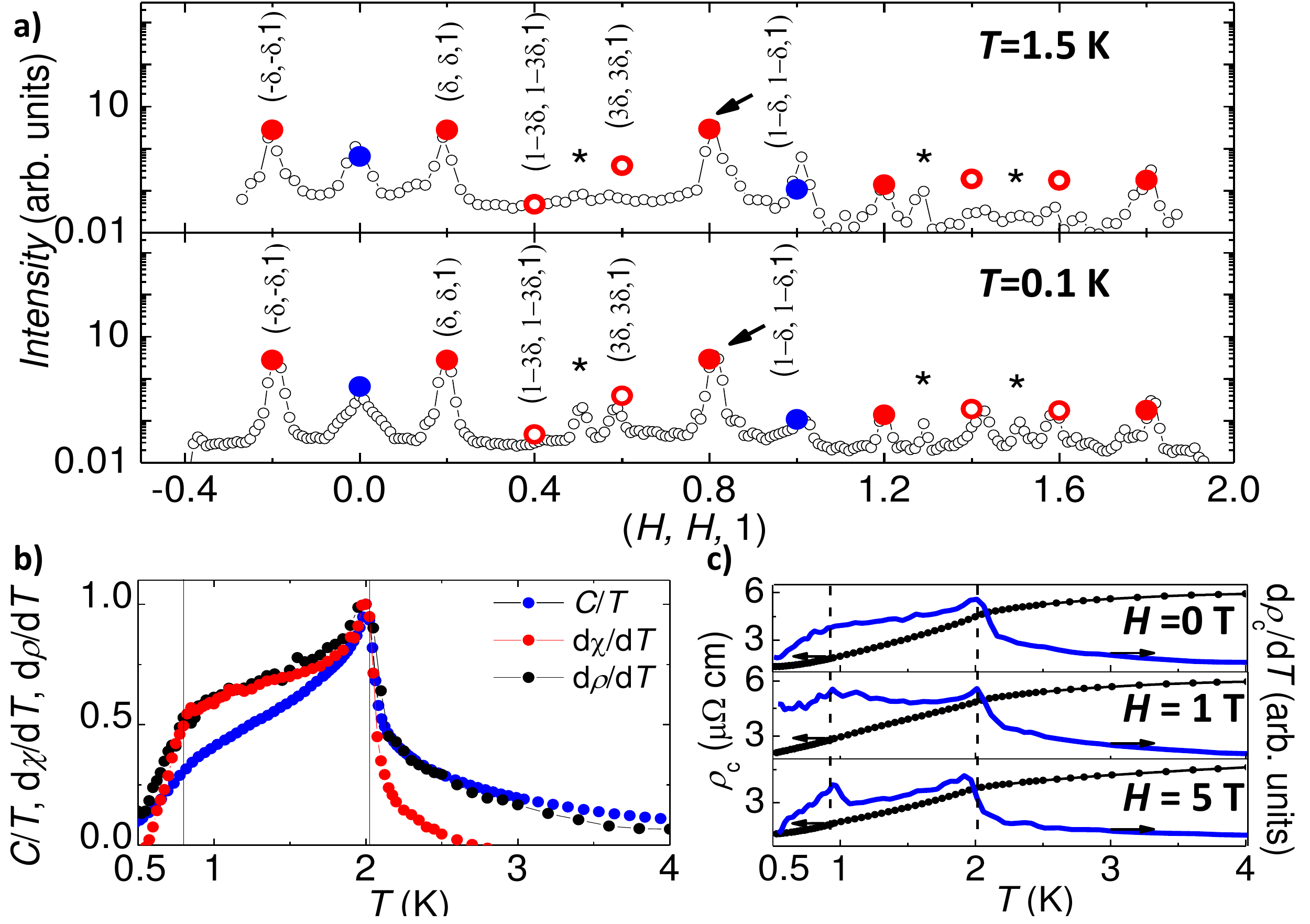}%
\caption{\label{fig_4} a) (HH1) scans at 1.5 K (top, CNCS) and 0.1 K (bottom, DCS). Blue circles are nuclear intensities, filled (open) red points are fundamental (first harmonic) AF peaks from our magnetic model, $\star$ marks Al peaks. b) Temperature dependencies of the specific heat C/T, and the temperature derivatives of $\rho$ and $\chi$, the latter measured in a dc field of 0.1 T along (110). All data are normalized at T$_{N}$=2.07 K to emphasize similar temperature dependencies. c) $\rho$(T) and $\partial\rho$/$\partial$T at different fields H$\|$(110). Vertical dashed lines in b) and c) are at T$_{N}$=2.07 K and 0.8 K. }
\end{figure}

The magnetic structure of \ypp~ is based on stripes of moment-bearing and non-moment bearing configurations of two classical Yb moments. It is reminiscent of the striped order found by nuclear magnetic resonance experiments in ~\scbo~ ~\cite{kodama2002,takigawa2013}, where increasing fields drive the SSL dimers based on Cu-based S=1/2 quantum spins through a sequence of discrete patterns of singlet and triplet states, reflecting the frustration of short-ranged exchange interactions inherent to the SSL~\cite{momoi2000,miyahara2000}. We find a much larger separation between the moment-bearing stripes in metallic \ypp, suggesting that here the frustration may involve long-ranged interactions, perhaps mediated by the conduction electrons. The low fields required for the suppression of AF order, for the cascade of magnetization steps that suggest intermediate structures ~\cite{shimura2012,iwakawa2012}, and for  SL saturation in ~\ypp~ may permit future investigations of the intermediate field states using neutron scattering, currently impossible in ~\scbo~ where fields approaching 30 T are required to drive analogous modulated phases~\cite{sebastian2008,jaime2012,matsuda2013}.

The strong Ising anisotropy that restricts the Yb moments to the (110) and (-110) directions has far-reaching implications for the AF order in ~\ypp. Unlike ~\scbo~ and also TmB$_{4}$~\cite{siemensmeyer2008}, where the moments are perpendicular to the SSL planes with a single AF wave vector, two perpendicular SLs are required in ~\ypp~ to accommodate the Ising moments \cite{ochiai2011}, which can be independently polarized by field. Once one SL is fully polarized, the SSL motif of orthogonal dimers is destroyed in favour of a square lattice of the remaining Yb dimers with spacing that approaches 8 $\AA$, although the same T$_{N}$=2.07 K that was found at H=0 when both SLs order is maintained. This is a very different AF state than has been envisaged for SSL systems with Heisenberg spins~\cite{shastry1981}, and it is possible that SSL physics does not act alone in determining the underlying magnetism of  ~\ypp.

We acknowledge valuable discussions with T. Sakakibara. Work at Stony Brook University was supported by NSF-DMR-1310008. Use of the NSLS was supported by the U.S. Department of Energy, Office of Science, Office of Basic Energy Sciences, under Contract No. DE-AC02-98CH10886. This work utilized facilities at NCNR that were supported in part by the National Science Foundation under Agreement No. DMR-0944772.  Research conducted at SNS was sponsored by the Scientific User Facilities Division, Office of Basic Energy Sciences, US Department of Energy.
%

\end{document}